\documentclass[12pt]{article}
\usepackage{epsfig}
\usepackage{subfigure}

\def\lbabar{\mbox{{\large\sl B}\hspace{-0.4em} {\normalsize\sl A}\hspace{-0.03em}{\large\sl B}\hspace{-0.4em} {\normalsize\sl A\hspace{-0.02em}R}}}
\usepackage{relsize}
\def\babar{\mbox{\slshape B\kern-0.1em{\smaller A}\kern-0.1em
    B\kern-0.1em{\smaller A\kern-0.2em R}}}

\def\pim   {\ensuremath{\pi^-}}

\def\Kbar  {\kern 0.2em\overline{\kern -0.2em K}{}}

\def\Kstarz  {\ensuremath{K^{*0}}}

\def\Kzb   {\ensuremath{\Kbar^0}}
\def\KzKzb {\ensuremath{K^0 \kern -0.16em \Kzb}}

\def\Dbar  {\kern 0.2em\overline{\kern -0.2em D}{}}

\def\Dzb   {\ensuremath{\Dbar^0}}
\def\DzDzb {\ensuremath{D^0 {\kern -0.16em \Dzb}}}
\def\Dstar   {\ensuremath{D^*}}

\def\Bz    {\ensuremath{B^0}}
\def\B     {\ensuremath{B}}
\def\Bbar  {\kern 0.18em\overline{\kern -0.18em B}{}}

\def\Bzb   {\ensuremath{\Bbar^0}}
\def\Bu    {\ensuremath{B^+}}
\def\Bub   {\ensuremath{B^-}}
\def\Bpm   {\ensuremath{B^\pm}}

\def\BB    {\ensuremath{B\Bbar}} 
\def\BzBzb {\ensuremath{B^0 {\kern -0.16em \Bzb}}}

\def\jpsi  {\ensuremath{{J\mskip -3mu/\mskip -2mu\psi\mskip 2mu}}} 
\def\psitwos {\ensuremath{\psi{(2S)}}}
\mathchardef\Upsilon="7107
\def\Y#1S{\ensuremath{\Upsilon{(#1S)}}}

\def\FourS {\Y4S}

\mathchardef\Deltares="7101
\mathchardef\Xi="7104
\mathchardef\Lambda="7103
\mathchardef\Sigma="7106
\mathchardef\Omega="710A
\def\Deltabar   {\kern 0.25em\overline{\kern -0.25em \Deltares}{}}
\def\Lbar {\kern 0.2em\overline{\kern -0.2em\Lambda\kern 0.05em}\kern-0.05em{}}
\def\Sigbar{\kern 0.2em\overline{\kern -0.2em \Sigma}{}}
\def\Xibar{\kern 0.2em\overline{\kern -0.2em \Xi}{}}
\def\Obar{\kern 0.2em\overline{\kern -0.2em \Omega}{}}
\def\Nbar{\kern 0.2em\overline{\kern -0.2em N}{}}
\def\Xbar{\kern 0.2em\overline{\kern -0.2em X}{}}

\def\mes        {\mbox{$m_{\rm ES}$}}

\def\ev   {\ensuremath{\rm \,e\kern -0.08em V}}
\def\kev  {\ensuremath{\rm \,ke\kern -0.08em V}} 
\def\mev  {\ensuremath{\rm \,Me\kern -0.08em V}} 
\def\gev  {\ensuremath{\rm \,Ge\kern -0.08em V}} 
\def\gevc {\ensuremath{{\rm \,Ge\kern -0.08em V\!/}c}} 
\def\tev  {\ensuremath{\rm \,Te\kern -0.08em V}}
\def\mevc {\ensuremath{{\rm \,Me\kern -0.08em V\!/}c}} 
\def\gevcc{\ensuremath{{\rm \,Ge\kern -0.08em V\!/}c^2}} 
\def\mevcc{\ensuremath{{\rm \,Me\kern -0.08em V\!/}c^2}}

\def\mum  {\ensuremath{\,\mu\rm m}} 

\def\invfb   {\ensuremath{\mbox{\,fb}^{-1}}}
\def\mus  {\ensuremath{\rm \,\mus}}

\def\ps   {\ensuremath{\rm \,ps}}

\def\mus        {\ensuremath{\,\mu{\rm s}}}    
\def\ps         {\ensuremath{{\rm \,ps}}}   
\def\gsim{{~\raise.15em\hbox{$>$}\kern-.85em
          \lower.35em\hbox{$\sim$}~}}
\def\lsim{{~\raise.15em\hbox{$<$}\kern-.85em
          \lower.35em\hbox{$\sim$}~}}

\def\CP                 {\ensuremath{C\!P}}

\def\to                 {\ensuremath{\rightarrow}}

\def\pep2{PEP-II}
\def\BF{$B$ Factory}

\def\mistag{\ensuremath{w}}

\def\deltaz{\ensuremath{{\rm \Delta}z}}
\def\deltat{\ensuremath{{\rm \Delta}t}}
\def\deltamd{\ensuremath{{\rm \Delta}m_d}}

\newcommand{\eqref}[1]{Eq.~(\ref{eq:#1})}

\newcommand{\epjc}      [1]  {{Eur.\ Phys.\ Jour.\ C~{\bf #1}}}

\def\jetset74   {\mbox{\tt Jetset \hspace{-0.5em}7.\hspace{-0.2em}4}}

\newcommand{\BABARPubYear}    {00}

\newcommand{\BABARProcNumber} {16}
\newcommand{\SLACPubNumber} {8625}

\newcommand{\LPNHENumber} {00-05}

\providecommand{\btodstarlnu}{\mbox{$B\to D^{*}\ell\nu$}}
\providecommand{\bztodstarlnu}{\mbox{$B^0\to D^{*-}\ell^+\nu$}}
\def\epem       {\ensuremath{e^+e^-}}

\begin{document}

\renewcommand{\thefootnote}{\fnsymbol{footnote}}

{\thispagestyle{empty}

\begin{flushright}
SLAC-PUB-\SLACPubNumber \\
\babar-PROC-\BABARPubYear/\BABARProcNumber \\
LPNHE-\LPNHENumber \\
September, 2000 \\

\end{flushright}

\par\vskip 2.5cm

\begin{center}
\Large \bf Measurements of \Bz~and \Bpm~lifetimes and \Bz-$\bar{\Bz}$
mixing with fully reconstructed \B~decays in \babar
\end{center}
\bigskip

\begin{center}
\normalsize
Fernando Mart\'{\i}nez-Vidal\\
LPHNE, IN2P3-CNRS/Universit\'es Paris 6 \& 7, France\\E-mail:
{\it martinef@SLAC.Stanford.EDU} \\
\large
(for the \lbabar\ Collaboration)

\end{center}
\bigskip \bigskip

\begin{center}
\large \bf Abstract
\end{center}

Time-dependent
\BzBzb\ flavor oscillations and \Bz\ and \Bu\ lifetimes 
are studied in a sample of fully
reconstructed \B\ mesons collected with the \babar\ detector, running
at the \pep2\ asymmetric \epem\ \BF\ with
center-of-mass energies near the \FourS\ resonance. 
This is the first time that time-dependent mixing and lifetime
measurements have been performed at \FourS\ energies.

\vfill
\begin{center}
\vskip1cm
\mbox{Contributed to the Proceedings of the 30$^{th}$ International 
Conference on High Energy Physics,} \\
7/27/2000--8/2/2000, Osaka, Japan
\end{center}

\vspace{1.0cm}
\begin{center}
{\em Stanford Linear Accelerator Center, Stanford University, 
Stanford, CA 94309} \\ \vspace{0.1cm}\hrule\vspace{0.1cm}
Work supported in part by Department of Energy contract DE-AC03-76SF00515.
\end{center}

\newpage



%
\pagestyle{plain}

\section{Introduction}

\mbox{Preliminary measurements of time-dependent} \BzBzb\ flavor 
oscillations and \Bz\ and \Bu\ lifetimes have been
performed with the \babar\ detector.  These analyses
exploit the copious production of \B\ meson pairs
in \FourS\ decays, produced by asymmetric \epem\ collisions
at the \pep2\ \BF\ at SLAC.
These measurements can be used to test  theoretical models of heavy
quark decay and to constrain the Unitarity Triangle (via the
sensitivity to the value of the Cabibbo-Kobayashi-Maskawa matrix \cite{KM} element $V_{td}$).
The data set, collected from January to June, 2000, has an 
integrated luminosity of 8.9 \invfb\ on the \FourS\ resonance and 
0.8 \invfb\ collected 40 \mev\ below the \BB\ threshold. This 
corresponds to about $(10.1\pm 0.4) \times 10^6$ produced \BB\ pairs.
The resolution function and mistag rates determined from data
in the analyses described here are also used in \CP\ asymmetry 
measurements \cite{BabarPub0001}. 

\section{Experimental Method}

The \babar\ detector is described in detail elsewhere
\cite{BabarPub0017}.  The analyses described here use
all the detector capabilities, including high resolution
tracking and calorimetry, particle identification and vertexing.

At \pep2\, the \B~meson pairs produced in the decay of the
\FourS\ resonance are moving in the lab frame along the
beam axis ($z$ direction) with a Lorentz boost of $\beta_z \gamma = 0.56$.
One \B~($\B_{\rm REC}$) is fully reconstructed in an all-hadronic
(\Bz $\to D^{(*)-} \pi^+,~D^{(*)-} \rho^+,~ D^{(*)-} a_1^+$,~\jpsi 
\Kstarz\ and \Bub $\to D^{(*)0} \pim,~ \jpsi K^-,~ \psitwos K^-$) or 
semileptonic decay mode ($\bztodstarlnu$) \footnote{Throughout this paper, 
conjugate modes are implied.}. 
A total of about 2600 neutral,
and a similar number of charged, \B\ candidates is
reconstructed in hadronic decay modes, with an average purity close to $90\%$. 
The background is mainly combinatorial. About 7500 \Bz's
are reconstructed in semileptonic modes, with an average purity of $\sim
84\%$. Backgrounds to the semileptonic mode are due to combinatorics, 
fake leptons, $c\bar c$ events, and charged \B\ decays from
$B^-\rightarrow D^{*+}(n\pi) l^- \nu$. Fig. \ref{fig:breco} shows the
beam-energy substituted \B\ 
mass (\mes) distributions for the hadronic sample (left)
and the $D^* - D^0$ mass distribution for the semileptonic sample (right)
\cite{BabarPub0007,BabarPub0008}.

\begin{figure}[htb]
\vspace{9pt}
\begin{center}
\begin{tabular}{cc}   
  \epsfig{file=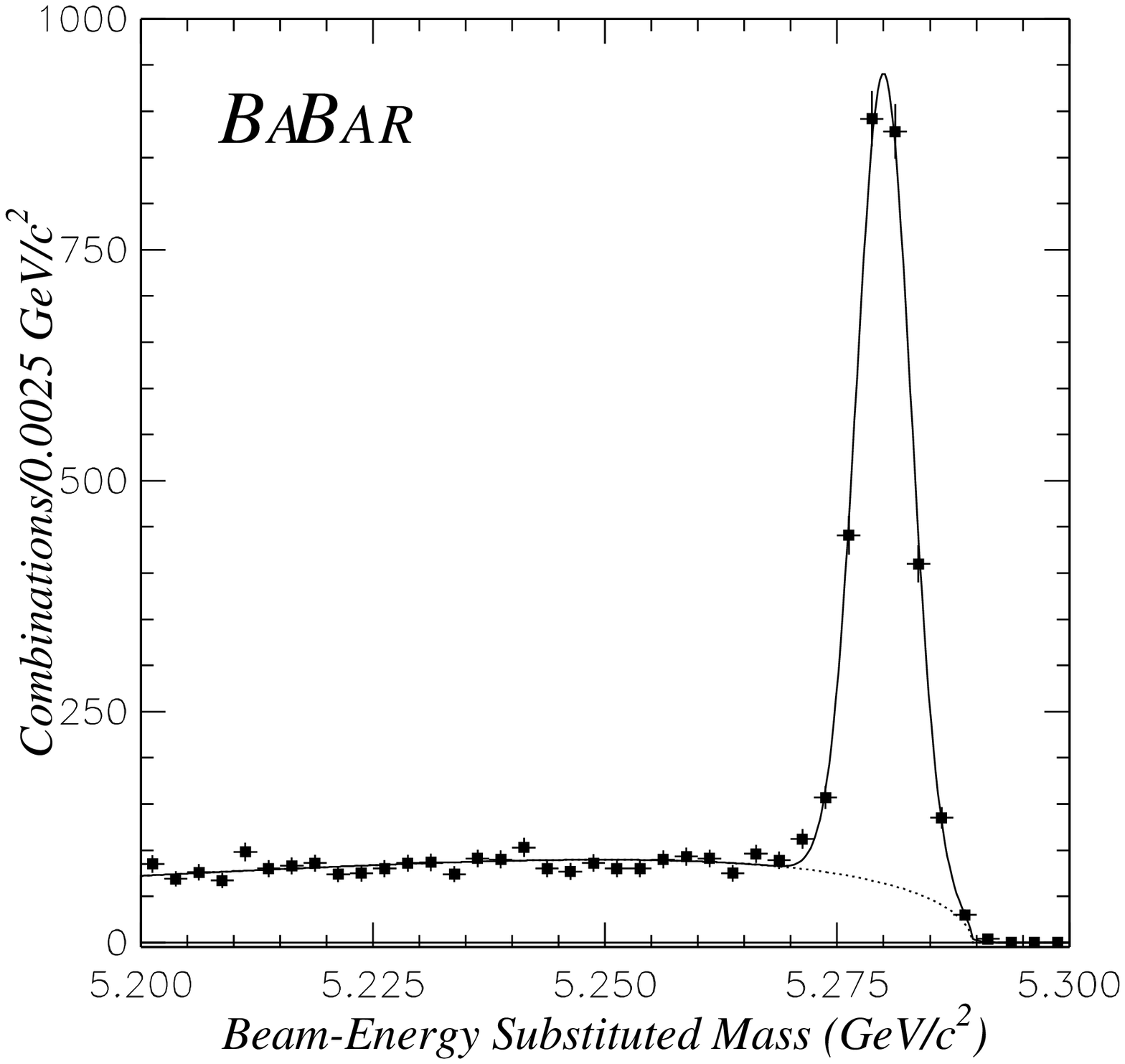,height=7.5cm}   & 
  \epsfig{file=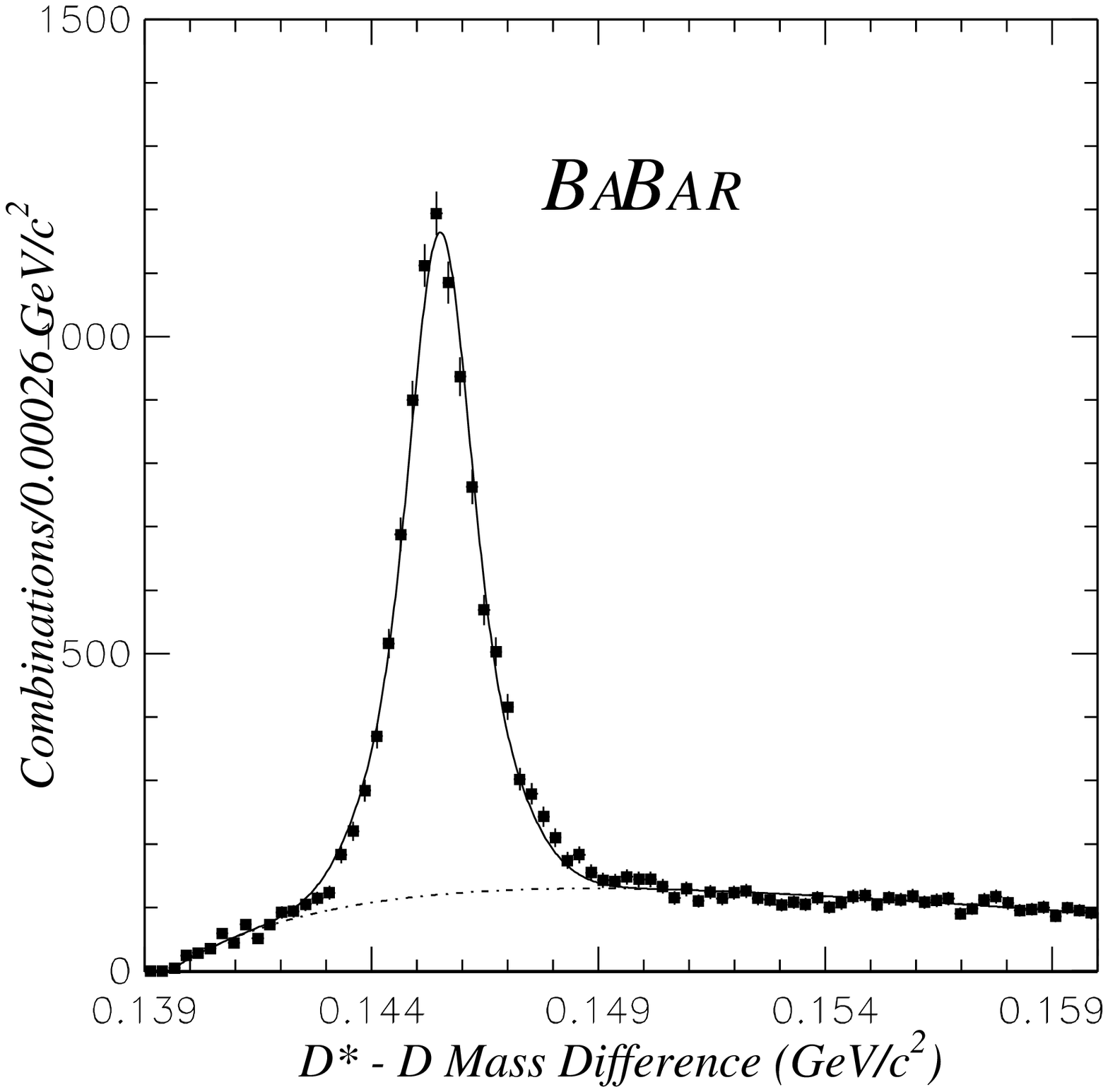,height=7.5cm}  \\
\end{tabular}
\caption{Left: Beam-energy substituted mass distribution (\mes) for all the
  hadronic \Bz\ modes. Right:  $D^*~-D^0$ mass difference distribution
  for the $\bztodstarlnu$ sample.}
\end{center}
\label{fig:breco}
\end{figure}

The separation between the two \B\ vertices along the boost direction,
$\Delta z = z_{\rm REC} - z_{\rm TAG}$, is measured and used to
estimate the decay time difference, $\Delta t \approx \Delta z/\beta_z
\gamma c$.
The  $\B_{\rm TAG}$ vertex is determined via
an inclusive procedure applied to all tracks 
not associated with the $\B_{\rm REC}$ 
meson \cite{BabarPub0007}. 
The typical separation between the two vertices is
$\deltaz = \beta_z \gamma c \tau_B \approx 260$\mum, to be compared to the
experimental resolution $\sim 110$\mum.
The $\Delta t$ resolution is dominated by the precision on the
$\B_{\rm TAG}$ vertex, and has little dependence on the decay mode of
the $\B_{\rm REC}$.  An event-by-event $\Delta z$ resolution is computed
and modified to fit the data by convolution with three Gaussians,
core, tail and outlier.
Most of the events, $\sim 70\%$, are in
the core Gaussian, with $\sigma \sim 0.6$ ps.

\section{Lifetime Measurements}

The \Bz\ and \Bu\ lifetimes are extracted from a simultaneous
unbinned maximum likelihood fit to the $\Delta t$ distributions of the
signal candidates, assuming a common resolution function.
Only hadronic decays from a subsample of 7.4 \invfb\ integrated
luminosity (on-resonance) have been used. 
An empirical description of the $\Delta t$ background shape is assumed, using \mes\
sidebands with independent
parameters for neutral and charged mesons. Fig. \ref{fig:lifetime}
shows the $\Delta t$ distributions with the fit result superimposed. Table \ref{tab:lifesyst}
summarizes the contributions to the total error (see \cite{BabarPub0007} for details).

\begin{figure}[htb]
\begin{center}
\vspace{9pt}
\begin{tabular}{cc}   
  \epsfig{file=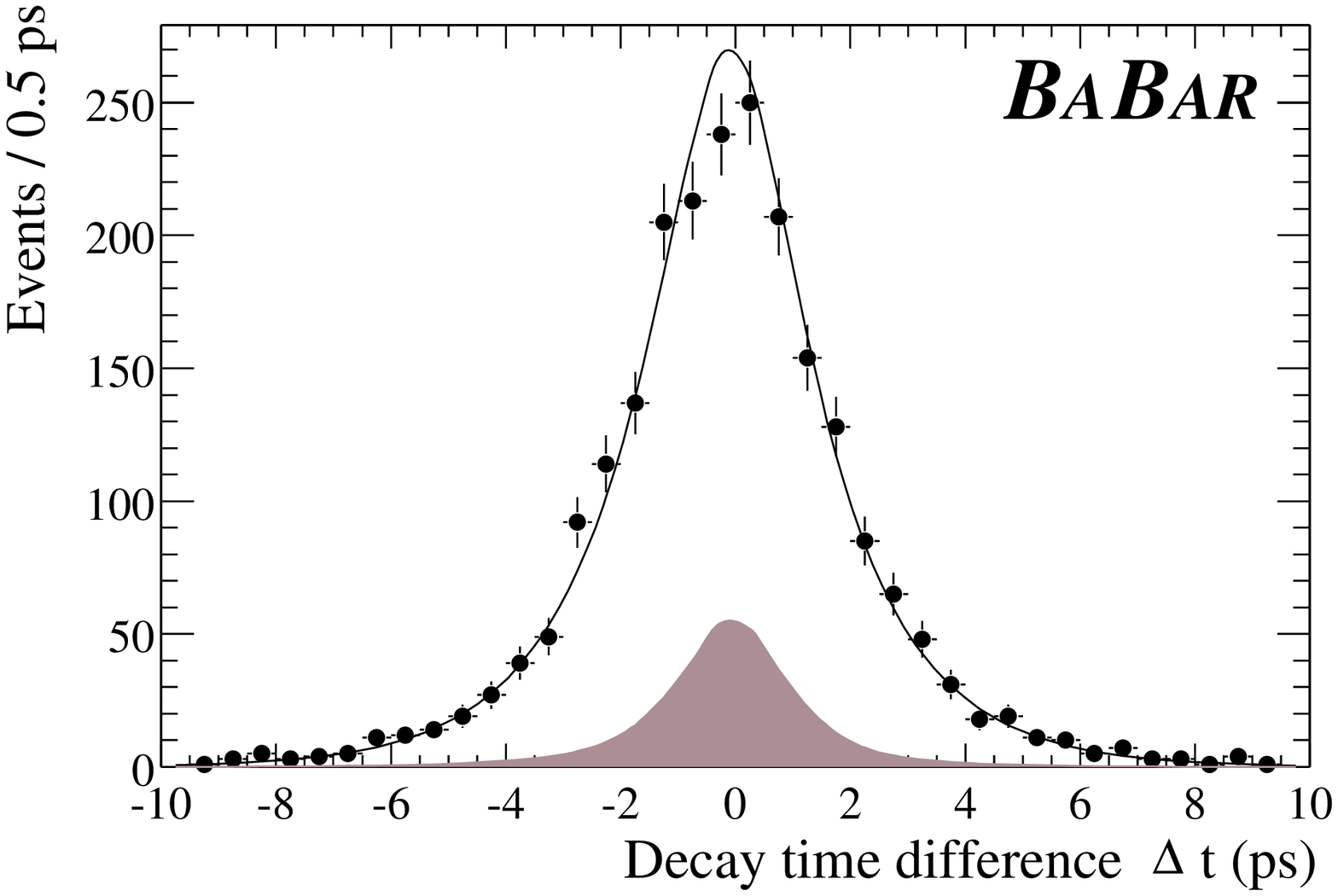,height=5.9cm}   & 
  \epsfig{file=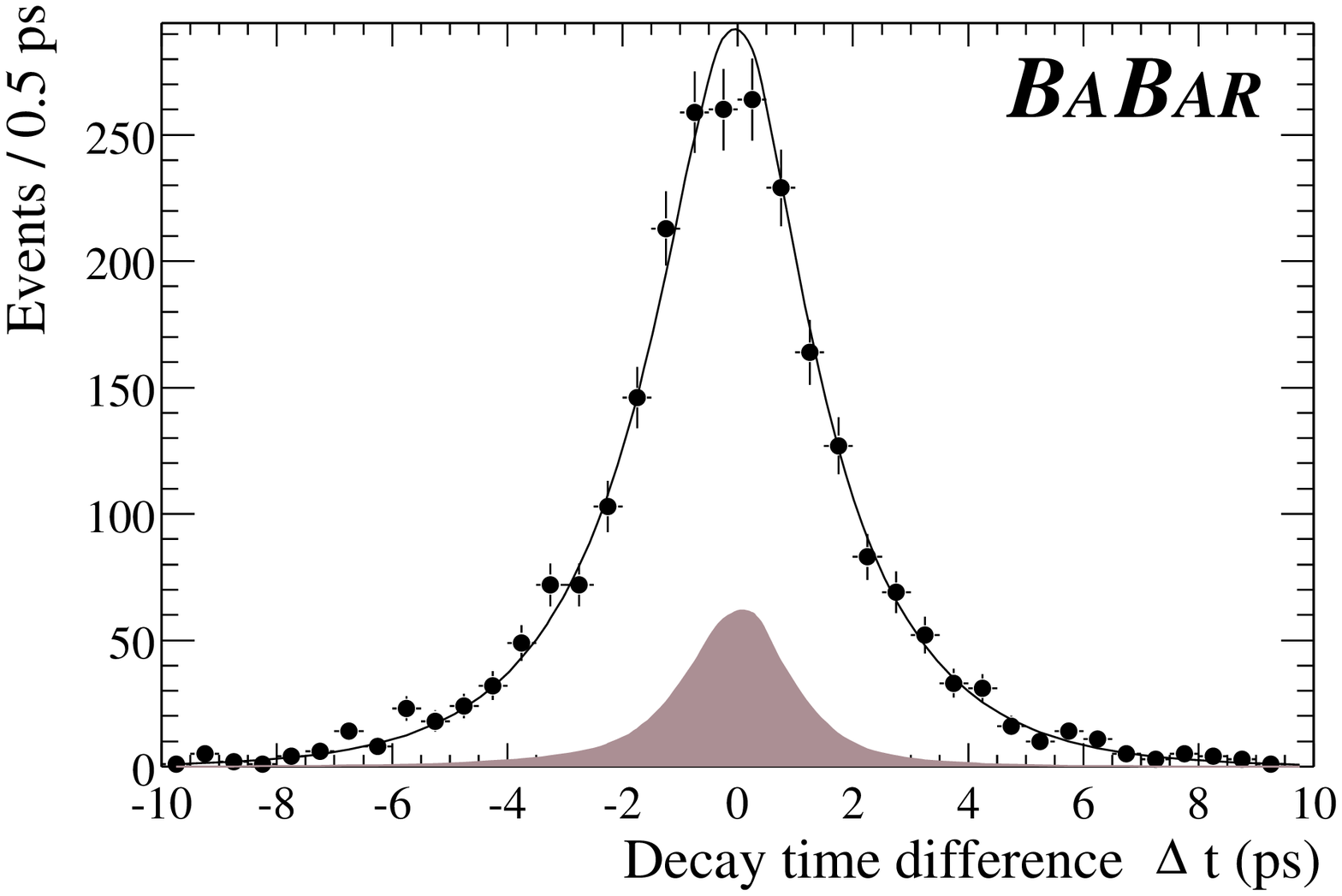,height=5.9cm}  \\
\end{tabular}
\caption{$\Delta t$ distributions for \Bz/\Bzb\ (left) and \Bu/\Bub\
  (right) candidates in the signal region (\mes$>5.27$ \gevcc). The result of the lifetime
fit is superimposed. The background is shown by the hatched area.}
\label{fig:lifetime}
\end{center}
\end{figure}

\begin{table}[htb]
\begin{center}
\caption{Summary of uncertainties for the \B\ Lifetime measurements.}\label{tab:lifesyst}
\begin{tabular}{|c|c|c|c|} 
\hline 
 
\raisebox{0pt}[6pt][2pt]{Source } & 
 
\raisebox{0pt}[6pt][2pt]{$\sigma(\tau_\Bz)$} & 
 
\raisebox{0pt}[6pt][2pt]{$\sigma(\tau_\Bu)$} &

\raisebox{0pt}[6pt][2pt]{$\sigma(\tau_\Bu / \tau_\Bz)$} \\

\hline
 
\raisebox{0pt}[6pt][2pt]{Data statistics} & 
 
\raisebox{0pt}[6pt][2pt]{0.052} & 

\raisebox{0pt}[6pt][2pt]{0.049} & 
  
\raisebox{0pt}[6pt][2pt]{0.044} \\

\hline

\raisebox{0pt}[6pt][2pt]{MC statistics} & 
 
\raisebox{0pt}[6pt][2pt]{0.016} & 

\raisebox{0pt}[6pt][2pt]{0.014} & 
  
\raisebox{0pt}[6pt][2pt]{0.014} \\

\raisebox{0pt}[6pt][2pt]{$\Delta t$ resolution} & 
 
\raisebox{0pt}[6pt][2pt]{0.007} & 

\raisebox{0pt}[6pt][2pt]{0.007} & 
  
\raisebox{0pt}[6pt][2pt]{0.008} \\

\raisebox{0pt}[6pt][2pt]{$\Delta z$ outliers} & 
 
\raisebox{0pt}[6pt][2pt]{0.016} & 

\raisebox{0pt}[6pt][2pt]{0.020} & 
  
\raisebox{0pt}[6pt][2pt]{0.005} \\

 

  

\raisebox{0pt}[6pt][2pt]{alignment} & 
 
\raisebox{0pt}[6pt][2pt]{-} & 

\raisebox{0pt}[6pt][2pt]{-} & 
  
\raisebox{0pt}[6pt][2pt]{-} \\

\raisebox{0pt}[6pt][2pt]{$z$ scale} & 
 
\raisebox{0pt}[6pt][2pt]{0.015} & 

\raisebox{0pt}[6pt][2pt]{0.016} & 
  
\raisebox{0pt}[6pt][2pt]{-} \\

\raisebox{0pt}[6pt][2pt]{boost} & 
 
\raisebox{0pt}[6pt][2pt]{0.006} & 

\raisebox{0pt}[6pt][2pt]{0.006} & 
  
\raisebox{0pt}[6pt][2pt]{-} \\

\raisebox{0pt}[6pt][2pt]{signal probability} & 
 
\raisebox{0pt}[6pt][2pt]{0.003} & 

\raisebox{0pt}[6pt][2pt]{0.002} & 
  
\raisebox{0pt}[6pt][2pt]{0.005} \\

\raisebox{0pt}[6pt][2pt]{background} & 
 
\raisebox{0pt}[6pt][2pt]{0.005} & 

\raisebox{0pt}[6pt][2pt]{0.017} & 
  
\raisebox{0pt}[6pt][2pt]{0.011} \\

\hline

\raisebox{0pt}[6pt][2pt]{Total systematics} & 
 
\raisebox{0pt}[6pt][2pt]{0.029} & 

\raisebox{0pt}[6pt][2pt]{0.035} & 
  
\raisebox{0pt}[6pt][2pt]{0.021} \\

\hline

\end{tabular}
\end{center}
\end{table}

\section{Time-dependent \BzBzb\ mixing}

A time-dependent \BzBzb\ mixing measurement 
requires the determination of the flavor of both \B's.
The $B_{\rm REC}$ flavor is known if it has been correctly reconstructed, 
and the flavor of the $B_{\rm TAG}$
is determined by 
exploiting the correlation between the flavor of the $\B_{\rm TAG}$
meson and the charge of its decay products \cite{BabarPub0008}. 
If there is an identified lepton its charge
is used; otherwise the summed charge of identified kaons provides the tag.
An event with no tagging leptons or kaons can still be tagged by
use of a neural net that exploits the flavor information carried by
other decay products, such as soft leptons from charm semileptonic decays and soft pions
from \Dstar\ decays.  

The effective flavor tagging efficiency is given by
$Q=\sum_i \epsilon_i (1-2\mistag_i)^2$ where 
the sum is over tagging categories, each characterized by 
a tagging efficiency $\epsilon_i$ and a probability to mis-identify 
the \B\ flavor, $\mistag_i$.  $Q$ is related to
the statistical significance of the measurement ($1/\sigma_{stat}^2
\sim N_{\B_{\rm TAG}} Q$).

From the time-dependent rate of mixed ($N_{mix}$) and unmixed
($N_{unmix}$) events, the mixing asymmetry 
$a(\Delta t) = (N_{unmix}-N_{mix})/(N_{unmix}+N_{mix})$ is
calculated as a function of $\Delta t$ and fit to the expected
cosine distribution, 
$$a(\Delta t) \propto (1-2\mistag) \cos \deltamd
\deltat \otimes {\cal {R}}( \Delta t | \hat{a} ) ,$$ 
where $\hat a$ are the parameters of the resolution function \cite{BabarPub0008}.
A simultaneous unbinned likelihood fit to all the tagging categories,
assuming a common resolution function,
allows the determination of both \deltamd\
and the mistag rates, $\mistag_i$.
An empirical description of the backgrounds is determined
by fitting to background control samples taken from data, allowing for
the following components: i) zero lifetime, ii) non-zero lifetime with no
mixing, iii) non-zero lifetime with mixing (only for semileptonic decays).
Fig. \ref{fig:mixing}
shows the $a(\Delta t)$ distributions with the fit result superimposed. Table \ref{tab:mixingsyst}
summarizes all the contributions to the total error (see
\cite{BabarPub0008} for details).

\begin{figure}[htb]
\vspace{9pt}
\begin{center}
\begin{tabular}{cc}   
  \epsfig{file=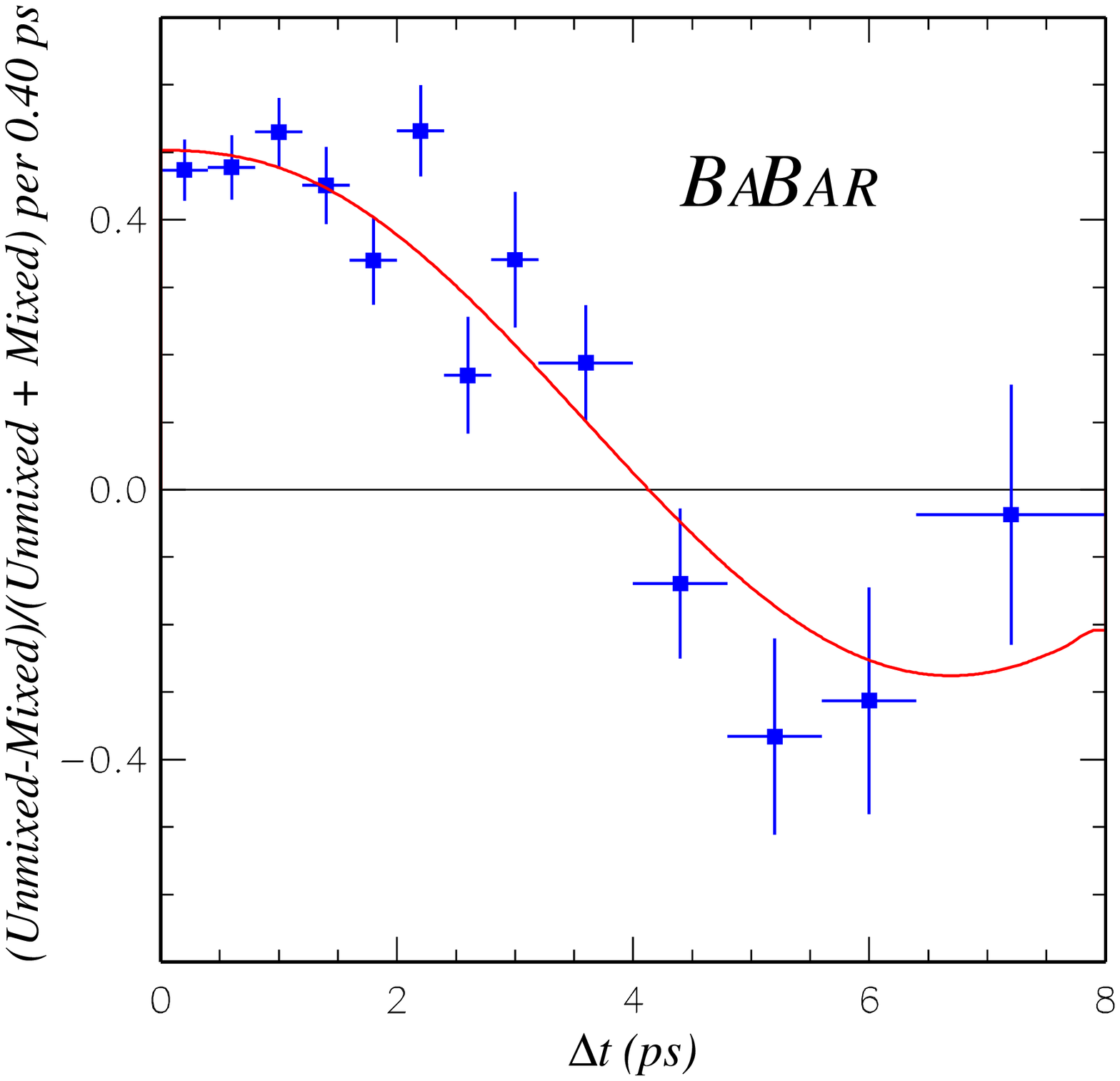,height=8.0cm}  &
  \epsfig{file=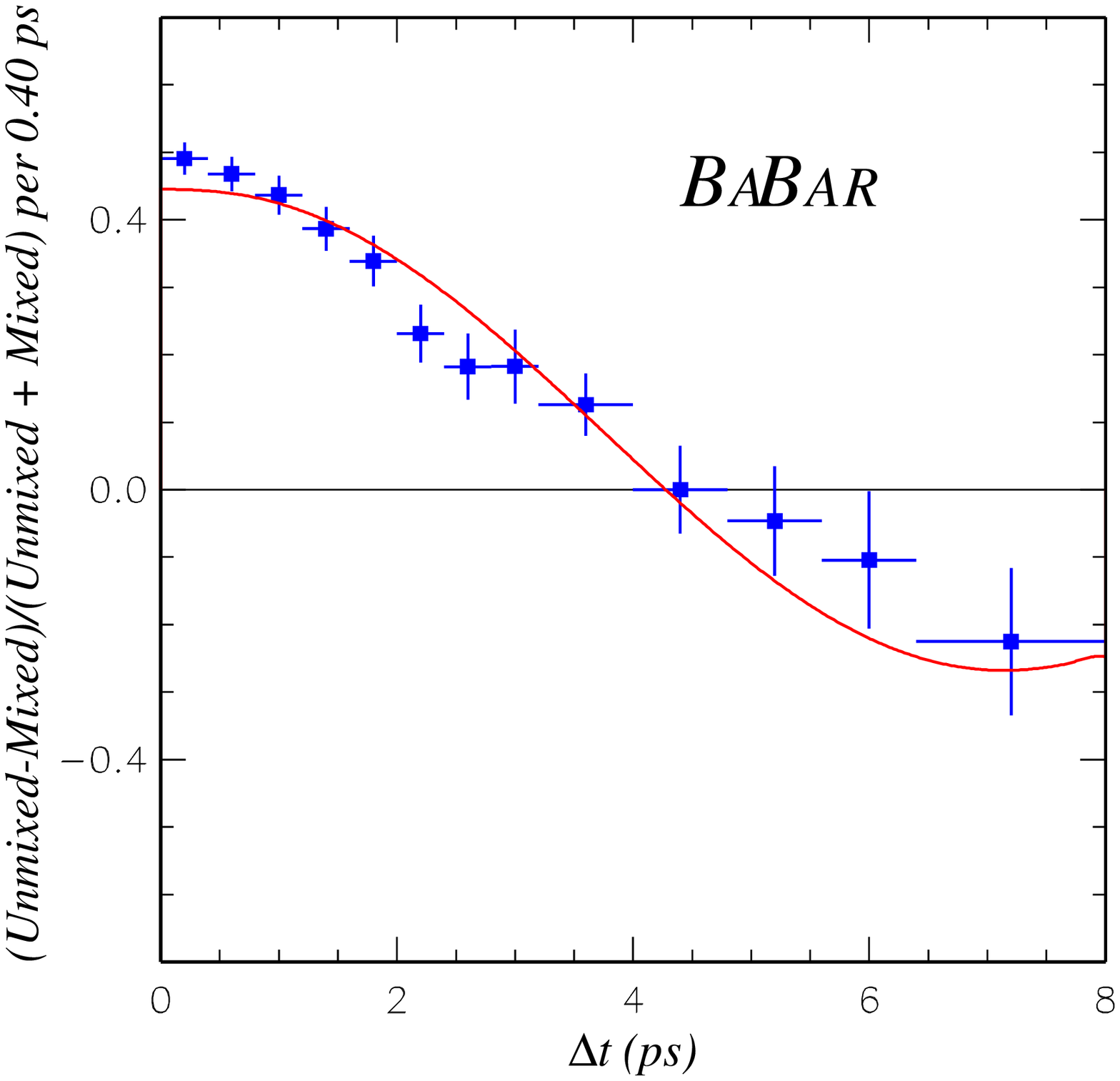,height=8.0cm}  \\
\end{tabular}
\caption{Time-dependent asymmetry $a(\Delta t)$ between unmixed and
      mixed events for (left) hadronic $B$ candidates with $\mes >
      5.27$\gevcc and (right) for \btodstarlnu\ candidates.}
\label{fig:mixing}
\end{center}
\end{figure}

\begin{table}[htb]
\begin{center}
\caption{Break-down of uncertainties for the \B\ Mixing measurements.}\label{tab:mixingsyst}
\begin{tabular}{|c|c|c|} 
\hline 
 
\raisebox{0pt}[6pt][2pt]{Source } & 
 
\raisebox{0pt}[6pt][2pt]{$\sigma(\deltamd)$} & 
 
\raisebox{0pt}[6pt][2pt]{$\sigma(\deltamd)$} \\

\raisebox{0pt}[6pt][2pt]{ } & 
 
\raisebox{0pt}[6pt][2pt]{hadronic} & 
 
\raisebox{0pt}[6pt][2pt]{semileptonic} \\

\hline
 
\raisebox{0pt}[6pt][2pt]{Data statistics} & 
 
\raisebox{0pt}[6pt][2pt]{0.031} & 

\raisebox{0pt}[6pt][2pt]{0.020} \\

\hline
 
\raisebox{0pt}[6pt][2pt]{MC statistics} & 
 
\raisebox{0pt}[6pt][2pt]{0.011} & 

\raisebox{0pt}[6pt][2pt]{0.009} \\

\raisebox{0pt}[6pt][2pt]{$\Delta t$ resolution} & 
 
\raisebox{0pt}[6pt][2pt]{0.011} & 
  
\raisebox{0pt}[6pt][2pt]{0.012} \\

\raisebox{0pt}[6pt][2pt]{background $\Delta t$} & 
 
\raisebox{0pt}[6pt][2pt]{0.002} & 
  
\raisebox{0pt}[6pt][2pt]{0.002} \\

\raisebox{0pt}[6pt][2pt]{background dilutions} & 
 
\raisebox{0pt}[6pt][2pt]{-} & 
  
\raisebox{0pt}[6pt][2pt]{0.006} \\

\raisebox{0pt}[6pt][2pt]{background fractions} & 
 
\raisebox{0pt}[6pt][2pt]{0.004} & 
  
\raisebox{0pt}[6pt][2pt]{0.006} \\

\raisebox{0pt}[6pt][2pt]{\Bu\ background} & 
 
\raisebox{0pt}[6pt][2pt]{-} & 
  
\raisebox{0pt}[6pt][2pt]{0.010} \\

\raisebox{0pt}[6pt][2pt]{\Bz\ Lifetime} & 
 
\raisebox{0pt}[6pt][2pt]{0.005} & 
  
\raisebox{0pt}[6pt][2pt]{0.006} \\

\raisebox{0pt}[6pt][2pt]{$z$ scale} & 
 
\raisebox{0pt}[6pt][2pt]{0.005} & 
  
\raisebox{0pt}[6pt][2pt]{0.005} \\

\raisebox{0pt}[6pt][2pt]{boost} & 
 
\raisebox{0pt}[6pt][2pt]{0.003} & 
  
\raisebox{0pt}[6pt][2pt]{0.003} \\

\hline

\raisebox{0pt}[6pt][2pt]{Total systematics} & 
 
\raisebox{0pt}[6pt][2pt]{0.018} & 
  
\raisebox{0pt}[6pt][2pt]{0.022} \\

\hline

\end{tabular}
\end{center}
\end{table}

\section{Results}

The preliminary results for the \B\ meson lifetimes are
\begin{eqnarray*}
\tau_{\Bz} &=& 1.506 \pm 0.052\ {\rm (stat)} \pm 0.029\ {\rm (syst)}\ \ps, \\ 
\tau_{\Bu} &=& 1.602 \pm 0.049\ {\rm (stat)} \pm 0.035\ {\rm (syst)}\ \ps
\end{eqnarray*}
and for their ratio is
$$\tau_{\Bu}/\tau_{\Bz} = 1.065 \pm 0.044\ {\rm (stat)}\ \pm 0.021\ {\rm (syst)}.$$

From the hadronic \Bz\ sample we measure the \BzBzb\ oscillation frequency:
\begin{eqnarray*}
  \Delta m_d  &=&  0.516 \pm 0.031\ ({\rm stat})  \pm 0.018  ({\rm
  syst})\  \hbar {\rm ps}^{-1}
\end{eqnarray*}

\noindent
and from the $D^{*-}\ell^+\nu$ sample the result is
\begin{eqnarray*}
  \Delta m_d    &=&   0.508 \pm 0.020\ ({\rm stat}) {}\pm 0.022 ({\rm
  syst})\ \hbar  {\rm ps}^{-1}.
\end{eqnarray*}

\noindent
Combining the two \deltamd\ results,
we obtain the preliminary result: 
\begin{eqnarray*}
\Delta m_d &=& 0.512 \pm  0.017 ({\rm stat})  \pm 0.022 ({\rm syst})\ \hbar {\rm ps}^{-1}.
\end{eqnarray*}
The mistag rates and $\Delta t$ resolution
function extracted from these fits are used 
in the \babar\ \CP\ violation asymmetry analysis \cite{BabarPub0001}.
The effective flavor tagging efficiency is found to be
$Q \approx 28\%$.

The above results are consistent with previous
measurements~\cite{PDG} and  are of
similar precision. They are also compatible with 
other \babar\ measurements \cite{christophe1,christophe2}. Significant
improvements are expected in the near future with the accumulation of
more data and further systematic studies.



 

\end{document}